# Deflection of ultra slow light under gravity

N. Kumar
Raman Research Institute, Bangalore 560080, India

Abstract

Recent experiments on ultra slow light in strongly dispersive media by several research groups reporting slowing down of the optical pulses down to speeds of a few metres per second encourage us to examine the intriguing possibility of detecting a deflection or fall of the ultra slow light under Earth's gravity, *i.e.,* on the laboratory length scale. In the absence of a usable general relativistic theory of light waves propagating in such a strongly dispersive optical medium in the presence of a gravitational field, we present a geometrical optics based derivation that combines '*the effective gravitational refractive index*' additively with the usual optical dispersion. It gives a deflection, or the vertical fall Δ for a horizontal traversal *L* as

$$\Delta = \frac{L^2}{2}\left(\frac{R_{\oplus G}}{R_{\oplus}^2}\right) n_g \left(\frac{1}{1+n_g \frac{R_{\oplus G}}{R_{\oplus}}}\right),$$

where $R_{\oplus G}/R_{\oplus}$ is the ratio of the gravitational Earth radius $R_{\oplus G}$ to its geometrical radius $R_{\oplus}$, and $n_g$ is the group refractive index of the strongly dispersive optical medium. The expression is essentailly that for the Newtonian fall of an object projected horizontally with the group speed $v_g = c/n_g$, and is tunable refractively thorugh the index $n_g$. For $L \sim 1$ m and $n_g = c/v_g \sim 10^8$ (corresponding to the ultra-slow pulse speed ~few x 1 ms$^{-1}$), we obtain a fall $\Delta \sim 1 \mu m$, that should be measurable – in particular through its sensitive dependence on the frequency that tunes $n_g$.

In a recent flurry of research publications [1-5] several research groups have reported spectacular slowing down of light pulses propagating in strongly dispersive media. Group velocities down to seven orders of magnitude smaller than the speed of light in vacuum have been observed. Such a slowing down is, of course, a direct consequence of the Kramers-Kronig (KK) dispersion relation between the real (reactive) and the imaginary (absorptive) parts of the frequency (ω) dependent dielectric constant ε(ω) = n(ω)$^2$. It gives a large group index $n_g = n + \omega \cdot \frac{\partial n}{\partial \omega}$ for a steeply rising refractive index (*n*) in the vicinity of a sharp absorption dip produced coherent optically through the interference of atomic transition amplitude as, *e.g.,* in the Electromagnetically Induced Transparency (EIT) [2]. Essential to using ultra slow light is the high group index $n_g(\omega) \gg n(\omega)$, but one with a low loss. This the EIT can provide, admittedly though in a rather narrow spectral window just a few Hz wide. While details of the physics informing the ultra slow light experiment are not without some controversy [6,7], its realizability in principle follows robustly from the KK relations. This has encouraged us to look into the intriguing possibility of there being a detectable fall of the ultra slow light pulse under Earth's gravity on a laboratory length-scale – indeed, in a "table-top" experiment. The phenomenon of fall of light under gravitation, commonly referred to as the bending or deflection of light (after the celebrated observation of the ray deflection during the 1919 total solar eclipse testing affirmatively Einstein's general relativity) necessarily involves astronomical length scales because of the enormity of the speed of light ($c = 3 \times 10^{10}$ cms$^{-1}$) in vacuum and the weakness of the stellar gravity $2GM/c^2 R \ll 1$. The ultra slow light at group speeds $v_g \ll c$ may thus provide the possibility for observing a sensible fall of light pulses in the strongly dispersive medium (with $n_g \sim 10^8$ or more, giving the group speed ~few metres per second or less) on the laboratory length scales under Earth's relatively much weaker gravity.

In this preliminary treatment of the ultra slow light propagating in a strongly dispersive optical medium in the presence of weak gravitational field, we make use of the idea of an "effective refractivity" ($n_g$) [8, 9] for the weak gravitational field, and add it to the group refractive index ($n_g$) for the dispersive medium. For a laboratory scale experiment with all reference length scales << the geometrical radius ($R_\oplus$) as well as the gravitational radius ($R_{\oplus G}$) of the Earth, we can write down the effective group refractive index $n(x,z)$ as

$$n(x,z) = n_g + n_G(x,z) \equiv n_g + \frac{2GM_\oplus}{(c/n_g)^2} \cdot \frac{1}{R_\oplus + z}, \qquad (1)$$

where the gravitational refractivity varies along the z axis in the vertical (x,z) plane. Here the replacement of $c$ by ($c/n_g$) on the right-hand side of Eq. (1) is justified by the following physically robust argument: In vacuum, the essentially exact GR deflection of light (zero rest mass photon) by a gravitating centre (e.g., the Sun of mass M) is $4GM/bc^2$, while the corresponding expression for the Newtonian gravitational deflection of a mass point coming in at speed $v$ and with the same impact parameter $b$ is $2GM/bv^2$, which is smaller by just a factor of 2 if we set $v = c$. While this celebrated asymptotic factor of 2 was crucial to the test of general relativity, it is not quite as important for an order of magnitude estimate of the fall of the ultra slow light on the laboratory scale under Earth's gravity. In fact, the optical pulse in the strongly dispersive medium moving at the ultra slow group speed $\ll c$ does suggest adopting the Newtonian view point, with the proviso that the light pulse too must fall under gravity following the equivalence principle of general relativity.

Thusly motivated, consider the trajectory of the ultra slow light pulse (wave packet) launched horizontally at $x=0$, $z=h$, along the x-axis in the dispersive optical medium described by the group refractive index in Eq. (1). The trajectory is then described in ray optics by [10]

$$\frac{d}{ds}\left(n(x,z)\frac{d\mathbf{r}}{ds}\right) = \nabla n(x,z), \qquad (2)$$

giving for the x-component

$$\frac{d}{ds}\left(n(x,z)\frac{dx}{ds}\right) = 0, \qquad (3)$$

inasmuch as the index is constant along the x-axis. Here s is the arc length along the ray trajectory.

Equation (3) is sufficient for our purpose. It says that $n(x,z) \cos\theta$ = constant along the trajectory, θ being the angle that the trajectory makes with the x-axis. More explicitly, we have

$$\frac{dz}{dx} = -\sqrt{\left(\frac{n^2(z)}{n^2(h)} - 1\right)} \qquad (4)$$

with $n(z) \equiv n(x,z)$ given by Eq. (1). Note that $n(x,z)$ is independent of x in our small length-scale limit.

Equation (4) is readily solved to give the fall Δ as function of the horizontal traversal L as

$$\Delta = \frac{L^2}{2}\left(\frac{R_{\oplus G}}{R_{\oplus}^2}\right) n_g \left(\frac{1}{1 + n_g \frac{R_{\oplus G}}{R_{\oplus}}}\right) \qquad (5).$$

Here we have used the obvious approximations

$$\Delta \ll R_{\oplus}, R_{\oplus G}.$$

Equation (5) is our main result. For a choice of parameters involved, $n_g = c/v_g$ ~$10^8$ (corresponding to the ultra slow light having speed ~1 $ms^{-1}$), and $L = 1\ m$, we obtain $\Delta$ ~ 1 μm. (Here we have used for the mass of the Earth $M_{\oplus}$ ~ $6 \times 10^{27}\ g$, and for its radius $R_{\oplus}$ ~ $6 \times 10^8\ cm$). The linear deflection Δ so obtained is a sensibly sized quantity. It can, however, be magnified, *e.g.*, by multiple passes. An interferometric detection should indeed be considered. More importantly, it can be discriminated against other factors by its tunability with frequency over a rather narrow range of 100 Hz or so in EIT. The frequency can be moreover modulated and the deflection $\Delta$ detected phase-sensitively.

Some remarks seem to be in order now. First, an appropriate generalization of the well known theorem, namely that light in vacuum follows a null geodesic, to the case of a dispersive optical (material) medium is called for. For a non-dispersive optical medium, there exists the optical metric of Gordon [10] whose null geodesics describe the geometrical optical light rays. A physical argument suggests that the refractive index *n* should then be replaced by the group inded $n_g$. In any case, an experimental test of the predicted fall of ultra slow light should be the real arbiter. Ideally, an isotropic homogeneous continuum such as an atomic gas/liquid should be used for this purpose as the nonlinear slow-wave optical medium under conditions of EIT. In such an optical medium, one expects no non-gravitational force to be exerted by the medium that would otherwise cause deviation from the geodesic of the optical metric.

Finally, there is a rather intriguing possibility where the ultra slow light is replaced by some other slow wave such as an ultra slow ultrasound. After all, even though sound is intrinsically mechanical involving material particulate motion (and unlike light, sound does not propagate in vacuum), the propagating sound wave is free and not bound to the material medium – it too should be subject to the equivalence principle and must fall under gravity. Again, in an isotropic medium one expects no non-gravitational deviation from a geodesic. Thus, a well collimated ultra slow ultrasound beam should be an interesting pesematological object of study – conceptually as well as experimentally.

Acknowledgment
The author would like to thank Andal Narayanan, Hema Ramachandran and Reji Philip for discussion on several points of EIT and the generation of ultra slow light. Thanks are also due to Anders Kastberg for his active interest in this work.